\documentclass[useAMS,usenatbib]{mn2e}
\usepackage{amsmath,amssymb,graphicx}
\def\rmmat#1{{\hbox{\rm #1}}}
\def\rmscr#1{\rmmat{\scriptsize #1}}

\newcommand{\comment}[1]{\relax}

\begin{document}

\title[Transit Timing]{Using 
long-term transit timing
to detect terrestrial planets}
\author[J. S. Heyl \& B. J.  Gladman]{Jeremy S. Heyl\thanks{email:
  heyl@phas.ubc.ca; Canada Research Chair}, Brett J. Gladman\thanks{Canada Research Chair}\\
Department of Physics and Astronomy, University of British Columbia, 
Vancouver, British Columbia, V6T 1Z1, Canada}

\date{Accepted ---.  Received ---; in original form --- }

\pagerange{\pageref{firstpage}--\pageref{lastpage}} \pubyear{2007}

\maketitle
\label{firstpage}

\begin{abstract}
  We propose that the presence of additional planets in extrasolar
  planetary systems can be detected by long-term transit timing
  studies.  If a transiting planet is on an eccentric orbit then the
  presence of another planet causes a secular advance of the
  transiting planet's pericenter over and above the effect of general
  relativity.  Although this secular effect is impractical to detect
  over a small number of orbits, it causes long-term differences in
  when future transits occur, much like the long-term decay observed
  in pulsars.  Measuring this transit-timing delay would thus allow
  the detection of either one or more additional planets in the system
  or the first measurements of non-zero oblateness ($J_2$) of the
  central stars.
\end{abstract}

\section{Introduction}
\label{sec:introduction}

The study of long-term orbital precession was one of the triumphs of
celestial mechanics, when the planetary theories of Laplace and
Lagrange showed that essentially all the known long-term precessions
of the planetary orbits could be explained by their mutual
gravitational interaction.  The perturbation caused by the small
planetary masses `breaks' the perfect central force character of the
Sun's gravitational field, causing the planetary orbital nodes to
regress and their perihelia to slowly advance, with typical periods of
$10^4$--$10^5$ years.  It was this advanced understanding of celestial
mechanics that permitted LeVerrier and Adams to detect a new planet in
our Solar System (Neptune) by inverting its observable effect on the
known planets to predict Neptune's mass and position.

The only known exception to these predictions at the start of the 20th century 
was the mystery that Mercury's perihelion longitude advanced
0.42\arcsec/yr faster than the predicted rate of $\simeq$5.31\arcsec /yr 
produced by the perturbative effects of all the other planets.  
The obvious possibility was that the Sun's mass
distribution was not spherically symmetric, but explaining the
mercurian advance would require a solar oblateness (measured by the
parameter $J_2$) of several percent, which was uncomfortably large
\citep[e.g.][]{1900AJ.....20..185H}.  General relativity provided the
solution, explaining essentially the entire discrepancy.  In fact the
Sun's oblateness is left with only an empirical upper bound of $J_2 <
1 \times 10^{-6} $, with a current theoretical estimate for its true
value of of $J_2 \sim 1\times10^{-7}$ \citep{2003Ap&SS.284.1159P}

As of the beginning of 2006, nearly two hundred extrasolar planets
have been discovered and several of them exhibit transits 
(as catalogued by \citealt{2006ApJ...646..505B} and references therein, especially
\citealt{2004A&A...415..391M}, \citealt{2005ApJ...632..638V}).
The longitude of periastron of eccentric hot Jupiters (3-day periods) will
show a secular advance of the instant of transit.  If the orbital
period can be well established (via transit timing or radial velocity)
then the long-term drift of the transit centers will allow one to
measure the slow advance of periastron.  Observations on 10-year
baselines should certainly show the relativistic advance, which is
much more important for close-in hot Jupiters since they are closer to
their parent stars than Mercury is to our Sun.  In this paper we
discuss the possibility of using the periastron advance rate to
measure either a host-star oblateness or the presence (and thus
discovery) of additional planets in the system.

\citet{2002ApJ...564.1019M} provided the first estimates of the
various contributions to the periastron advance of extrasolar planets.
\citet{2005MNRAS.359..567A} and \cite{2005Sci...307.1288H} build upon
this work by including the effects of resonances and including other
contributions to the timing noise of planetary transits.  Although
these numerical studies automatically include the secular advance of
periastron, they focus more on the stociastic variation of the
interval between transits.  This work as \citet{2002ApJ...564.1019M}
did focusses specifically on the secular periastron advance and builds
upon that work by including a more accurate calculation of the advance
for planets whose orbits have similar semimajor axes and by outlining
several techniques to measure the periastron advance and their
associated precision (including the effects of general relativity). 

In \S\ref{sec:secular-advance} we calculate the secular advance of
periastron of the orbit of a planet around a star due to other planets
in the system under reasonable approximations.  We also present the
relativistic and quadrupole contributions to the periastron advance.
\S\ref{sec:what-sens-other} places these calculations in the context
of extrasolar planetary systems. \S\ref{sec:how-can-one} presents
analytic and numerical estimates of how well we can determine the
advance of periastron using various techniques (timing of the primary
and secondary transits with or without radial velocity information).
The special relativistic corrections are outlined in
\S\ref{sec:spec-relat-corr}, and \S\ref{sec:outlook} outlines the
prospects of this technique using the planets discovered so far as a
guide.

\section{Secular advance}
\label{sec:secular-advance}

In the Newtonian two-body problem the Laplace Runge-Lenz vector
(also known as Hamilton's vector, or the eccentricity vector),
which points from the star to the planetary orbit's pericenter, is
stationary, 
so the location of periastron is constant.
Several effects can cause the periastron to advance.
In our Solar system in increasing order of importance we have :
\begin{enumerate}
\item
Stellar oblateness,
\item
General relativity, and
\item
Other planets.
\end{enumerate}
The amplitude of these effects depends on several parameters, and thus
in extrasolar planetary systems the order of importance may differ. 
In the following sections we provide expressions for the rate of periastron
advance for these effects.

\subsection{Stellar contributions}
\label{sec:stell-contr}
The central star can cause periastron advance by either being non-spherical
or due to general relativistic effects caused by its mass.
These effects cause a periastron advance \citep{Misn73} of
\begin{equation}
\delta \varpi = \frac{6\pi G M_*}{c^2 a(1-e^2)} + J_2 \frac{3\pi R_*^2}{a^2 (1-e^2)^2}
\end{equation}
per {\it radial} period due to the star itself.  Here, $c$ is the speed of
light, $G$ Newton's gravitational constant, $a$ the semimajor axis of
the planet's orbit, $e$ its eccentricity, and $M_*$, $R_*$, and $J_2$
are the mass, radius, and oblateness parameter of the star,
respectively.

Since $a = (P^2 GM_*/4\pi^2)^{1/3}$,
we get a time rate of change for the relativistic component of
\begin{equation}
{\dot \varpi_\rmscr{GR}} =\frac{\delta \varpi}{P} = \frac{121 ''{\rm~yr}^{-1}}{1-e^2} \left ( \frac{M_*}{M_\odot} \right )^{2/3} \left ( \frac{P}{3 {\rm ~day}}\right )^{-5/3}
\label{eq:varpiGR}
\end{equation}
and for the stellar oblateness
\begin{equation}
{\dot \varpi_{J_2}} = \frac{3.1 ''{\rm ~yr}^{-1}}{(1-e^2)^2} \frac{J_2}{10^{-6}} \left ( \frac{R_*}{R_\odot} \right )^{2} \left ( \frac{M_*}{M_\odot} \right )^{-2/3} \left ( \frac{P}{3 {\rm ~day}}\right )^{-7/3} 
\label{eq:J2}
\end{equation}
where we have scaled these effects to typical values appropriate for a
hot Jupiter.
For Mercury the relativistic effect is 0.43\arcsec /yr.
This drops as $a^{-5/2}$ for more distant orbits, being
0.086\arcsec /yr for Venus, 0.038\arcsec /yr for Earth, and 0.013\arcsec /yr for Mars.  
The scaling value of $J_2=10^{-6}$ used in Eq.~\ref{eq:J2} is essentially 
a firm upper limit for the oblateness of our sun that can be obtain 
from theoretical and observational considerations \citep{2003Ap&SS.284.1159P}.

\subsection{Planetary contributions}
\label{sec:plan-contr}
Just as Mercury's pericenter advance is affected by the other planets, an
exoplanet's orbit will precess due to the perturbations of an unseen
planet.  We will generally assume that a second planet in the system
is an external one, although the theory is almost identical if the
perturber is interior.

A simple way to estimate the contribution to the periastron advance
from another planet in the system 
is to assume that the mass of the second planet is smeared out over
a circular ring coplanar with the first planet's orbit and calculate the
contribution of this ring to the potential.
If the orbit of the second planet is elliptical or not coplanar with
the observed planet, the eccentricity, inclination and longitude of
the ascending node will also change on a secular timescale. 

An additional planet in the system causes the periastron 
of an observed planet to advance by
\begin{equation}
\delta \varpi 
\approx
\frac{\pi}{2} \frac{m_2}{M_*} \frac{a^3 (3 a_2^2 - a^2)}{a_2  (a_2^2 - a^2)^2} 
 \approx \frac{3\pi}{2} \frac{m_2}{M_*} \frac{a^3}{a_2^3}
\end{equation}
during each radial orbit 
\citep{1979AmJPh..47..531P}.
The approximation holds if $a\ll a_2$ and
$a_2$ is the radius of the second planet's orbit.  The mass of second
planet is $m_2$.  

\begin{figure}
\includegraphics[width=3.4in]{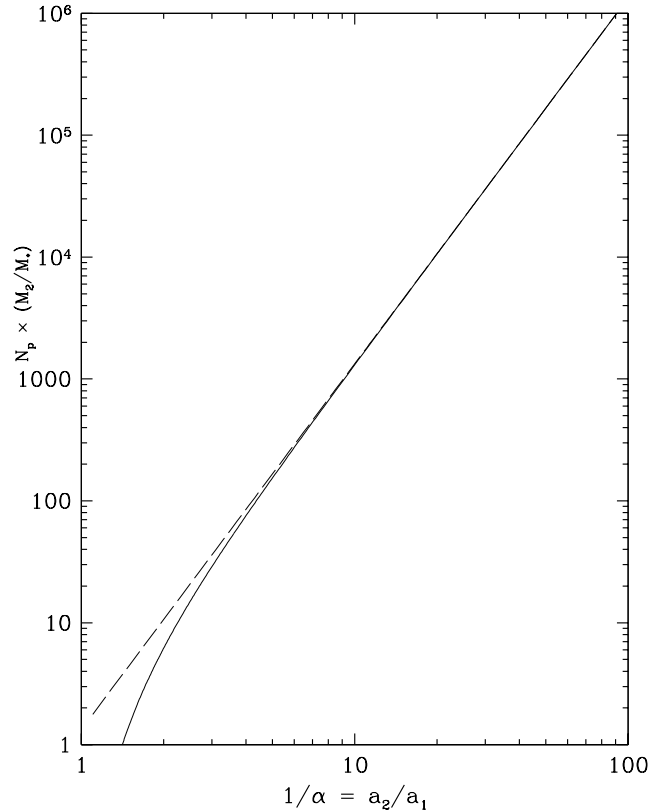}
\caption{The number of orbital periods of an inner planet required
for that planet's orbit to precess by $2\pi$ under the secular advance
induced by an external perturber, times the mass ratio of the perturber
to the central star ($m_2/M_*$).
The $x$-axis is the semimajor axis ratio of the outer to inner planet.
For an unseen Jovian perturber the number of orbits for precession
is the vertical axis time $10^3$ and for a 3 Earth-mass perturber one
multiplies by $10^5$.
The dashed line is the approximation using the small-$\alpha$ limit of
the Laplace coefficient.
\label{fig:Np}
}
\end{figure}

We can do a bit better if we relax the assumption of $a\ll a_2$.  
In this case we find that the potential due to the second planet is
\begin{equation}
V_2 = -\frac{G m_2}{a_2} \frac{2}{\pi \left (\alpha + 1 \right )} K \left ( \frac{2 \alpha^{1/2}}{\alpha+1} \right ) 
\end{equation}
where $K$ is the complete elliptic integral of the first kind and
$\alpha \equiv a/a_2$.
From \citet{1979AmJPh..47..531P}, the advance of periastron per radial orbit  
is
\begin{equation}
\delta \varpi = 2\psi - 2\pi\,
\end{equation}
where the increase in azimuthal angle for half a radial orbit is
\begin{equation}
\psi = \pi \left \{ 3 + a \left [ V''(a) / V'(a) \right ] \right \}^{-1/2}
\end{equation}
and $V(a)$ is the potential due to the star and the perturbing planet.   
The limit $V(a)=1/a$ for no perturbing planet yields $\psi=\pi$ as expected.
For $m_2\ll M_*$ we have
\begin{eqnarray}
\delta \varpi &=& \frac{m_2}{M_*} \frac{\alpha}{(\alpha+1)(\alpha-1)^2}
  \Biggr [ \left ( \alpha^2 + 1 \right ) E \left ( \frac{2
      \alpha^{1/2}}{\alpha+1} \right )  \nonumber \\ 
  & & ~~~-  \left ( \alpha - 1 \right )^2 K \left ( \frac{2
      \alpha^{1/2}}{\alpha+1} \right ) \Biggr ] \\
  &=& \frac{\pi}{2} \frac{m_2}{M_*} \alpha^2 b^{(1)}_{3/2}(\alpha) \rmmat{~for~} \alpha<1
\end{eqnarray}
where $E$ is the complete elliptic integral of the second kind.
The function $b^{(1)}_{3/2}(\alpha)$ is a Laplace coefficient from classical
perturbation theory; it is equal to $3\alpha$ for $\alpha\ll 1$,
is order unity for $\alpha$=0.1--0.5, and then increases
rapidly to $>$100 for $\alpha > 0.9$.  In the limit of a large ratio
between the two semimajor axes we obtain
\begin{eqnarray}
\delta \varpi  
&\approx& \frac{m_2}{M_*} \times \left \{ \begin{array}{ll} 
\frac{3\pi}{2} \alpha^3 & \alpha \ll 1, ~~ \mathrm{exterior~perturber} \\
\frac{3\pi}{2} \alpha^{-2} & \alpha \gg 1, ~~ \mathrm{interior~perturber} 
\end{array}
\right .
\end{eqnarray}
This is perhaps more transparently expressed by noticing that 
$N_p = 2\pi/\delta \varpi$ is the number of inner planet revolutions
required for a full precession of its orbit : 
\begin{equation}
N_p = \frac{2\pi}{\delta \varpi } = 
\frac{4}{\alpha^{2}\; b^{(1)}_{3/2}(\alpha)} \frac{M_*}{m_2}
\end{equation}
\noindent
for $a_1<a_2$.
Fig.~\ref{fig:Np} gives the exact value of $N_p \times (m_2/M_*)$
and its asymptotic form.
As examples, an unseen Jupiter-mass planet ($m_2/M_* \sim10^{-3}$) at ten times the 
semimajor axis of an interior hot-Jupiter (3-day period) will cause the hot
Jupiter's orbit to precess completely in about $10^6$ orbits = $3\times10^6$ days,
or about 8,000 years; a 3-Earth mass planet ($m_2/M_* \sim 10^{-5}$) 1.5 times more 
distant than a hot Jupiter would cause precession in only about $10^5$ orbits 
(800 years).

The analysis above greatly underestimates the effect if resonant or
near-resonant terms are important to the dynamics.  It is unclear
whether resonant configurations are ubiquitous or happenstance; some
extrasolar planetary systems are already known to exhibit
near-resonant behaviour
\citep{1992Natur.355..325R,2005Natur.434..873F}.
\citet{2005MNRAS.359..567A} and \citet{2005Sci...307.1288H} include
the effects of orbital resonances and the eccentricity of the
perturber's orbit; however, \citet{2005Sci...307.1288H} focus on the
stochastic variation of the inter-transit interval and
\citet{2005MNRAS.359..567A} do not consider secular terms in their
calculations, which is the focus of the analysis here.

An explicit expression for the precession rate of the longitude of
periastron, if the semimajor axis of the outer planet is much larger
than that of the inner observed planet ($a\ll\ a_2$), is
\begin{equation}
{\dot \varpi} = 355 ''{\rm~yr}^{-1} 
\frac{m_2}{M_\oplus} \frac{M_\odot}{M_*} 
\left ( \frac{a}{a_2} \right )^3 \frac{3 {\rm ~day}}{P} 
\end{equation}
where $P$ is the period of the inner planet.  This asymptotic
estimate of the rate of periastron advance (shown in Fig.~\ref{fig:Np}) 
significantly underestimates the rate for
$\alpha>1/4$.  In fact for $\alpha=0.5$ the estimate
falls short by a factor of two \citep[c.f.][]{2002ApJ...564.1019M}.


\section{What is the sensitivity to other planets in the system?}
\label{sec:what-sens-other}
\begin{figure}
\includegraphics[width=3.4in]{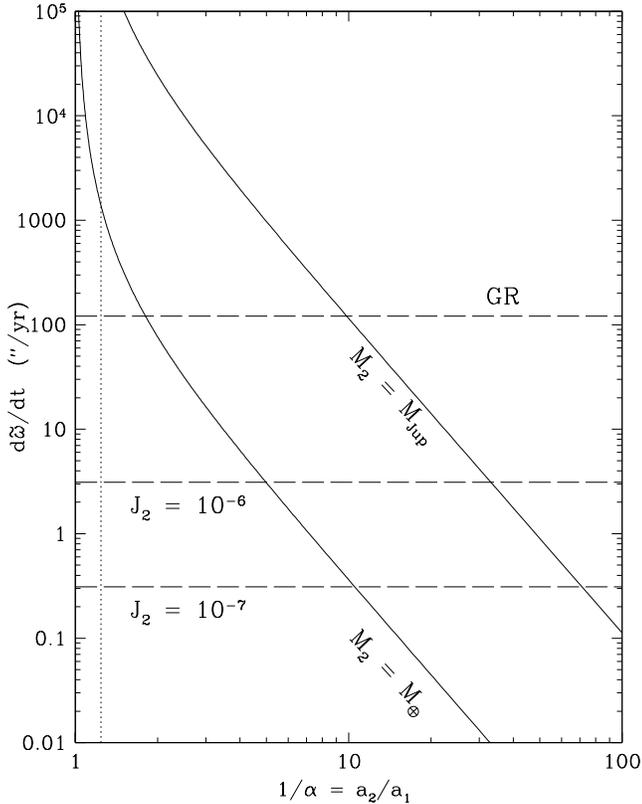}
\caption{The periastron precession rate of a hot Jupiter in a
  low-eccentricity, three-day orbit around a solar-mass star.  The
  vertical dotted line shows the minimum separation of an exterior
  Jupiter-mass planet (see text).  }
\label{fig:wdot}
\end{figure}

Fig.~\ref{fig:wdot} shows the precession contribution from the three
effects on an observed transiting planet. The most interesting
component is the contribution from other planets in the system, so we
would like to have firm upper limits on the contributions from the
other two effects in order to estimate the excess that might be due
to an unseen planet.

The relativistic contribution is the most certain.  The mass of the
star can generally be estimated to a few percent and the period of the
orbit of the planet can be measured to a part in 10,000 or better;
therefore, the value in Eq.~\ref{eq:varpiGR} can be estimated to a
high degree of precision.

Since the GR-induced precession rate is simply proportional to the
mass of the star, the fractional precision of this mass estimate will
set the sensitivity limit of the periastron advance method.  If we
take 3\% as a nominal $M_*$ precision then, if one observes a
transiting hot-Jupiter with $P$=3~days, a precession rate different
from the GR-rate by less than 3\% will not provide a reliable
detection of a planet or an oblate star; the 3-day period used for 
Fig.~\ref{fig:wdot} gives by chance that the induced rate from the
host star with $J_2=10^{-6}$ falls close to the detection limit.

The value of the oblateness ($J_2$) for the host stars of extrasolar
planets is unknown -- we don't even know it well for our Sun.
However, we can estimate an upper limit to its value from the solar
estimates and a scaling of the rotation rate of our sun to that of the
host star if it is known.  From observations of solar oscillations and
theoretical considerations, the value of $J_2$ for the sun is probably
around a few times $10^{-7}$ \citep{2003Ap&SS.284.1159P}.
\citet{2005ApJ...631.1215W} recently determined the spin rate of the
star in the transiting planetary system HD 209458.  The value that
they obtain $v \sin i = (4.70 \pm 0.16)\,$km/s is not much larger that
the typical values for the Sun of $1.4-2.0\,$km/s, so it is unlikely
that the value of $J_2$ for at least this system is much larger than
that of the Sun.  The value $10^{-6}$ probably provides an upper limit;
this yields a oblateness contribution of about 3" per year, an
order of magnitude less that the contribution from a nearby Earth-mass
planet.  It is unlikely that stars have values of $J_2>10^{-6}$;
detected precession rates more than $\sim$3\arcsec\ faster than the
GR-induced rate would be strong evidence for the presence of an unseen
planet.

The detection of a periastron advance rate for an extrasolar planet with
a three-day period that exceeded of the relativistic amount by 31" per
year (Eq.~(\ref{eq:J2}) would either indicate the presence of an
unseen planet or a stellar $J_2$ of about $10^{-5}$, an order of
magnitude larger than the upper limits for the Sun.  Either result is
interesting.  Assuming that the oblate star hypothesis is ruled out,
such a detection either implies (Fig.~\ref{fig:wdot}) a
terrestrial-scale planet a few times further than the hot Jupiter, or
a jovian-mass planet even more distant (for a given precession rate
there is a one-parameter family of mass-distance for the external
perturber).

The jovian-mass case would likely be limited by several constraints.
Very large precession rates (many times the GR-induced precession) are
unlikely to be found; if both planets were jovian mass, then the
analysis of \citet{Glad93} shows that an outer planet would not be
stable if $(a_2-a_1)/a_1 \lesssim 0.24$ ({\it i.e.,} the outer
planet's semimajor axis must be than $>$125\% of the inner planet's or
the system would be unstable; the vertical dotted line of
Fig.~\ref{fig:wdot}).  Notice that a jovian perturber an order of
magnitude further away than the hot Jupiter induces a precession rate
comparable to the GR-induced rate.  However, such a perturber should
be trivially detectable already in the radial velocity data from the
system; thus unless the host star is so noisy that radial-velocity
techniques cannot be applied, the periastron technique is less
sensitive to massive planets.

However, while terrestrial-mass planets cannot be seen by the
radial-velocity technique (the radial velocity signal is less than one
meter per second for periods greater than three days), the periastron
advance they would produce on a hot Jupiter is in principle detectable
if they are up to $\simeq5$ times more distant. 

\section{How can one observe the advance of periastron?}
\label{sec:how-can-one}

We can have the following information (loosely in order of difficulty)
\begin{enumerate}
\item
Timing of the primary transit,
\item
Timing of the secondary transit and
\item
Timing of radial velocity data 
\end{enumerate}

\subsection{Primary transit}
\label{sec:primary-transit}

As the orbit precesses the location of the transit relative to the
orbit changes; specifically, the value of the true anomaly ($\nu$) at
the center of the transit decreases at a rate of ${\dot \varpi} =
\delta \varpi/P$.  To get the time of the transit, we have to relate
the true anomaly $\nu$ to the mean anomaly (${\mathcal M}$ --- the phase of
the orbit increasing linearly in time from zero at one periastron to
$2\pi$ at the next) through Kepler's equation,
\begin{equation}
{\mathcal M} = {\mathcal E} - e \sin {\mathcal E},
\end{equation}
with
\begin{equation}
\tan \frac{\mathcal E}{2} = \sqrt{ \frac{1-e}{1+e} } \tan \frac{\nu}{2}
\end{equation}
where ${\mathcal E}$ is the eccentric anomaly.

For simplicity we work in the frame rotating with the precessing
orbit, so we take $P$ to be the time between successive 
periastrons.
The true anomaly that corresponds to the transit thus changes as the
orbit precesses. The interval between two transits is simply the
difference in the mean anomalies at each transit divided by the mean
motion $2\pi/P$.  If the orbit is not precessing, this time is simply
the period of the orbit.  However, if the periastron advances, the
timing between the transits is
\begin{equation}
\Delta t = P \left ( 1 - \frac{d{\mathcal M}}{d\nu} \frac{\delta \varpi}{2\pi} \right ) + {\mathcal O} (\delta\varpi^2)
\label{eq:1}
\end{equation}
where we have assumed that the periastron advance per orbit is small.
As the orbit precesses, the time between
transits will change, according to
\begin{eqnarray}
\frac{d \Delta t}{d t} &=& -\left ( 1 + \frac{d{\mathcal M}}{d\nu} \frac{\delta
  \varpi}{2\pi} \right )^{-2}  \frac{(\delta \varpi)^2}{2\pi}
\frac{d^2 {\mathcal M}}{d \nu^2}  \\
&=& - \frac{d^2 {\mathcal M}}{d \nu^2} \frac{(\delta \varpi)^2}{2\pi}  + {\mathcal O} (\delta\varpi^3)
\label{eq:2}
\end{eqnarray}
At face value this gradual change of the timing of the transits
appears hopeless to detect because the change is proportional to the
very small square of the advance of periastron per orbit.  However,
the difference in the timing accumulates from orbit to orbit as the
orbit precesses, so in practice one would predict the timing of the
transits from a few observations and look for a difference from that
prediction after many hundreds of orbits had passed.  Essentially we
are interested in the integral of $\Delta t$ over many periods.

If one observes several initial transits and determines a value of $\Delta
t$, one can predict the timing of future transits.  These predictions
will be incorrect for a precessing orbit by the following amount
\begin{equation}
t_{\rm pred} - t_{\rm actual} = \left ( {\mathcal M}(\nu) - {\mathcal M}(\nu_0) - \left . \frac{d {\mathcal M}}{d \nu} \right |_0 \left ( \nu - \nu_0 \right ) \right ) \frac{P}{2\pi}.
\end{equation}
where $\nu_0$ is the true anomaly during the transit at the initial
epoch and $\nu$ is the true anomaly during the transit at the later
epoch.

One can calculate
\begin{equation}
\frac{d^2 {\mathcal M}}{d \nu^2} = e \frac{ \sqrt{1-e^2} \sin \nu
  \left ( 1 - e \cos {\mathcal E} \right )  -(1-e^2)\sin {\mathcal E} }{\left ( 1 + e \cos \nu \right )^2}
\end{equation}
or expanding for small eccentricities
\begin{equation}
\frac{d^2 {\mathcal M}}{d \nu^2} = 2 e \sin \nu - 3 e^2 \sin 2\nu + 3 e^3 \sin 3\nu + {\mathcal O}(e^4)
\end{equation}
From Eq.~\ref{eq:2} one can see that for an error in the predictions
to accumulate the second derivative the mean anomaly with respect to
the true anomaly must not vanish. Thus, how quickly the error
accumulates depends on the initial epoch of the observations.  If the
transit is initially occurring at periastron ($\nu=0$) or apastron
($\nu=\pi$), the second derivative vanishes so it will take
significantly longer for the time delay to become observable.

To lowest order in the change in the periastron advance (or the true
anomaly at transit), we have
\begin{eqnarray}
t_{\rm pred} - t_{\rm actual} &\approx& \frac{P}{4\pi} \left ( \nu - \nu_0
\right )^2 \left . \frac{d^2 {\mathcal M}}{d \nu^2} \right |_0 \\ 
&\approx & 
\frac{P}{4\pi} \left [ \dot \varpi \left ( t - t_0 \right ) \right ]^2 \left . \frac{d^2 {\mathcal M}}{d \nu^2} \right |_0,
\label{eq:6}
\end{eqnarray}
so the delay accumulates quadratically in time.  Using reasonable
values for the various numbers we have
\begin{equation}
t_{\rm pred} - t_{\rm actual} 
= 1~{\rm ms} \frac{e\sin \nu_0}{0.1} \frac{P}{3~{\rm days}} 
\left ( \frac{t - t_0}{1~{\rm year}} \frac{\dot \varpi}{100''~{\rm yr}^{-1}}\right )^2 
\label{eq:3}
\end{equation}

\subsubsection{Error Analysis}
\label{sec:error-analysis}

Eq.~\ref{eq:3} makes it seem seem hopeless to detect the timing delay
because one can determine the time of a particular transit to possibly
ten seconds; therefore, naively one would expect to have to wait
one hundred years before detecting 
an advance with $t_{\rm pred} - t_{\rm actual} =$10~seconds.
Fortunately, one can
detect the periastron advance in the series of transit times long
before one could detect it in the timing of an individual transit.

In practice one characterizes the timing of the transits with a formula of the
following form
\begin{equation}
t_n = A + B n + C n^2 = t_0 + \Delta t_0 n + \frac{P}{4\pi} \left ( \delta \varpi  \right )^2 \left . \frac{d^2 {\mathcal M}}{d \nu^2} \right |_0 n^2
\end{equation}
where $n$ is the number of the transit.  $t_0$ is the time of
an initial reference transit, $\Delta t_0$ is the initial time between transits
and the quadratic term contains the periastron advance.

Using the standard results for $\chi^2$ fitting, we obtain
\begin{eqnarray}
\sigma_C &=& \sigma_0 \left (  \frac{180 r_0^4}{(r_0N)[(r_0N)^4-5(r_0N)^2+4]}
\right )^{1/2} \\ 
&\approx&  13.41 \sigma_0 r_0^{-1/2} N^{-5/2} \left ( 1  + {\mathcal O}(N^{-2}) \right ) 
\end{eqnarray}
where $N$ is the number of the last transit sampled, $r_0$ is the
fraction of transits with times and $\sigma_0$ is the timing error on
each transit.

The upper limit obtained for the value of the advance per orbit $\delta \varpi$ 
will be 
\begin{equation}
\sigma_{\delta \varpi} = \frac{\sigma_C^{1/2}}{2} \left ( \frac{P}{4\pi} \left . \frac{d^2 {\mathcal M}}{d \nu^2} \right |_0 \right )^{-1/2}
\end{equation}
where we have ignored the fractional error in the values of $P$ and
${d^2 {\mathcal M}}/{d \nu^2}$.  This yields
\begin{equation}
\sigma_{\delta \varpi} \approx 1.6 \times 10^{-5} \left (\frac{N}{1000}\right )^{-5/4} r_0^{-1/4} \left ( \frac{\sigma_0}{10~{\rm s}} \frac{3~{\rm day}}{P} \frac{0.1 |\sin \nu_0|}{e} \right )^{1/2}  
\end{equation}
as an upper limit on the advance per orbit.  This is not much larger
than the expected relativistic contribution of
\begin{equation}
\left ( \delta \varpi \right
)_{\rm GR} = 5 \times 10^{-6} \frac{1}{1-e^2} \left ( \frac{M_*}{M_{\rm
sun}} \right )^{2/3} \left ( \frac{P}{3~{\rm day}} \right )^{-2/3} 
\end{equation}
so for $N$ greater than a few thousand the relativistic term will
dominate over the statistical errors in the timing.  These estimates
agree with the results of \citet{2002ApJ...564.1019M} who considered
the effects of periastron advance on the timing of the primary transit
and the duration of the primary transit.

One thousand transits of a planet with a three-day orbit takes just a
shade under eight and a quarter years.  The upper limit on $\delta
\varpi$ decreases with time as $t^{-5/4}$ until a reliable periastron
advance is detected.  After this time, the errors on this detection
decrease as $t^{-5/2}$.  The time to achieve the desired sensitivity
scales as the $P^{3/5}$ so this technique is also applicable to
planets with longer orbital periods.  These estimates assume that
every transit is timed ($r_0=1$).  The error analysis assumes that the
observed transits are evenly spaced in time; it should be possible to
devise an observing strategy that achieves errors similar to the
$r_0=1$ case with many fewer observations --- this is beyond the scope
of this paper.

\subsubsection{Why don't observations of the primary transit tell us more?}
\label{sec:why-dont-observ}

Each time the planet orbits the star is takes a bit less than an
orbital period for it to reach the point of primary transit, because
the orbit is shifting a bit.  However, since we don't know the radial
orbital period itself, this time is essentially unobserved.  The time
for the planet to cover the missing angular distance is related to the
distance from the star to the planet at transit.  As the orbit
precesses, this distance will change which in turn will change the
time between transits.  It is this change in the time between transits
that we try to observe.  The correction in the time between transits
is proportional to the periastron advance.  The change in the distance
between the star and the planet from orbit to orbit is also
proportional to the periastron advance.  Combining these facts
indicates that the change in the time between transits is second-order
in the small periastron advance; consequently, it takes a relatively
many orbits to detect the periastron shift, if one times only one type
of transit.

\subsection{Secondary transit}
\label{sec:secondary-transit}

Looking at the secondary transit (when the planet goes behind the
star) does not just provide a new set of times to fit but also
provides new information and possibly a faster way of detecting unseen
planets in the system.  The secondary transit occurs when the true
anomaly is 180 degrees away from where the primary transit occurs;
therefore, we will denote quantities that describe the secondary
transit with the subscript $_\pi$.  The primary transit is given by
the subscript $_0$.  Let us examine at the time between two
successive primary and two successive secondary transits.  Using the
earlier formulae  we have to lowest order in the advance of
periastron
\begin{equation}
\Delta t_0 = P \left ( 1 - \left . \frac{d{\mathcal M}}{d\nu} \right |_0 \frac{\delta \varpi}{2\pi} \right )
\end{equation}
and
\begin{equation}
\Delta t_\pi = P \left ( 1 - \left . \frac{d{\mathcal M}}{d\nu} \right |_\pi \frac{\delta \varpi}{2\pi} \right ).
\end{equation}
where
\begin{eqnarray}
\frac{d{\mathcal M}}{d\nu} \!\! &=& \!\! \sqrt{1-e^2} \frac{1 - e \cos
  {\mathcal E} }{1 + e \cos \nu}
\nonumber \\
& = & 
1 - 2 e \cos \nu + \frac{3}{2} e^2 \cos 2 \nu - e^3 \cos 3 \nu + {\mathcal O}(e^4)
\end{eqnarray}
If we take the difference between these two values we get
\begin{equation}
\Delta t_0 - \Delta t_\pi =  \left ( \left . \frac{d{\mathcal M}}{d\nu} \right |_0 -\left . \frac{d{\mathcal M}}{d\nu} \right |_\pi \right ) \delta \varpi \frac{P}{2\pi}
\end{equation}
so the interval between two successive primary and two successive secondary transits differs by an amount proportional to the advance of periastron per orbit.  This should be compared with observations of the primary transit alone in which the advance of periastron only enters at second order.

Furthermore, the difference in the time between the primary and secondary transits also contains some valuable information.  We have
\begin{equation}
t_\pi - t_0 = \left [ {\mathcal M}(\nu_0+\pi) - {\mathcal M}(\nu_0) \right ] \frac{P}{2\pi}.
\end{equation}
If the orbit is eccentric this will differ from half of the orbital period.  Expanding in the eccentricity we have
\begin{equation}
t_\pi - t_0 = \frac{P}{2} + \frac{P}{2 \pi} \left [ 4 e \sin \nu_0 + \frac{2}{3} e^3 \sin 3 \nu_0 + {\mathcal O} ( e^5 ) \right ].
\end{equation}
The first term in the series is twice the value of the first term in
the series for $d^2 {\mathcal M}/d \nu^2$, so the interval between the
primary and secondary transits helps to calculate the periastron
advance when one uses the timing of the primary transits.

\subsubsection{Error Analysis}
\label{sec:error-analysis-1}

How well can we determine the time of the transits and the time
between successive transits?  Fitting
the transit times to a timing model
\begin{equation}
t_{0,n} = t_0 + \Delta t_0 n
\end{equation}
and
\begin{equation}
t_{\pi,n} = t_\pi + \Delta t_\pi n .
\end{equation}
Because the periastron advance now enters in the difference between
the interval between the successive transits, we only need to fit the
times to first order in the number of the transit ``$n$''.  From the
$\chi^2-$analysis we obtain the following error estimates
\begin{equation}
\sigma_{t_\pi} = \sigma_\pi  \left ( \frac{2 [2 (r_\pi N) + 1]}{(r_\pi N) [(r_\pi N)-1]} \right )^{1/2}
\end{equation}
and 
\begin{equation}
\sigma_{\Delta t_\pi} = \sigma_\pi  \left ( \frac{12 r_\pi^2}{(r_\pi N) [(r_\pi N)^2-1]} \right )^{1/2}
\end{equation}
where $r_\pi$ and $\sigma_\pi$ are the fraction of secondary transits
with times and the error in the timing of the secondary transit. The
error in the time of the initial transit scales as $N^{-1/2}$ where
$N$ is the number of orbits that have elapsed between the first and
last one observed.

We are interested in the differences
\begin{equation}
\Delta t_0 - \Delta t_\pi = \left ( -4 e \cos \nu_0 - 2 e^3 \cos 3 \nu_0  + {\mathcal O}(e^5) \right ) \delta \varpi \frac{P}{2\pi}
\end{equation}
and 
\begin{equation}
\sigma_{\Delta t_0-\Delta t_\pi} \approx \left [ \left ( \frac{\sigma_0^2}{r_0} + \frac{\sigma_\pi^2}{r_\pi} \right ) \frac{12}{N^3} \right ]^{1/2}
\end{equation}
for large $N$, and especially the error in
\begin{eqnarray}
\sigma_{\delta \varpi} &\approx& 7 \times 10^{-8} \frac{0.1}{e |\cos
  \nu_0|} \left ( \frac{N}{1000} \right )^{-3/2} \frac{3~{\rm day}}{P}
\times \nonumber \\
& & ~~~
\left ( \frac{\sigma_0^2}{(10~{\rm s})^2 r_0} + \frac{\sigma_\pi^2}{(10~{\rm s})^2 r_\pi} \right )^{1/2}.
\end{eqnarray}
Combining results for the secondary transit with those from the
primary transit yields an increase in sensitivity of a factor of 
$\sim200$.
If we could time the secondary transit to the same precision of ten
seconds it would take only $N\sim 100$ to detect an Earth-like planet
within twice the semimajor axis of the observed planet.  
With current instruments and the brightest targets, the secondary
transits can be timed to a precision of about 100~s (J. Matthews,
{\em priv. comm.}), yielding an estimate of about $N\sim 400$ orbits for
a similar detection.
The time to achieve the desired sensitivity scales as $P^{1/3}$ so
this technique is also applicable to planets with longer orbital
periods; furthermore, one is sensitive to smaller planets in systems
with larger eccentricities.

\subsubsection{Why does the secondary transit help so much?}
\label{sec:why-does-secondary}

After analyzing the primary transit, one saw how difficult it was to
disentangle the change in the angle of periastron from the
observations of the orbital period of the system.  The timing of the
secondary transit breaks this degeneracy, and it is straightforward to
understand why.  Unless the orbit is perfectly circular (or if we are
so unlucky as to have $\nu_0 \approx \pi/2$ or $3 \pi/2$ in the epoch
of observations), the planet is at different distance from the star at
the primary and secondary transit, so according to Kepler's Second Law
(conservation of angular momentum) its angular velocity along the
orbit is different at these two times.  If the periastron shifts as
the planet orbits, it takes a different amount of time to cover the
missing angle at the primary than at the secondary transit;
consequently, the time between secondary transits differs from that
between the primary transits --- if one can detect this time difference
one can detect the advance of periastron.

\subsection{Radial Velocity Information}
\label{sec:radi-veloc-inform}

The radial velocity information is arguably the most difficult to
obtain.  It turns out that it is essentially the least useful (at
least in quantity) for the purposes of characterizing the periastron
shift.  It is difficult to imagine obtaining timing of the radial
velocity data with a precision of tens of seconds, so it is not
directly useful in getting additional timing points, as we did with the
secondary transit. In principle, one would find that the time interval
between when the star passed through a particular radial velocity and
when it repeated itself would depend on the radial velocity in
question.

However, the {\it period} found by fitting the radial velocity curve is
typically precise to about one second .  This time interval would only
differ from the interval between transits by a tiny amount, on the
order of the periastron advance.  Determining accurately the
relationship between the time between periastrons (what we have called
the period) and the period found by fitting the radial velocities
requires Monte Carlo simulations of the observed data.  One can also
gain some insight into what time interval emerges from fitting radial
velocities by considering orbits that are nearly circular.

When one fits the radial velocity measurements one is most sensitive
to parts of the orbit with large accelerations to or from the
observer.  The acceleration along the line of sight reaches an extreme
when the jerk vanishes.  To first order in the orbital eccentricity
\begin{eqnarray}
\frac{d^2 v_{\rm los}}{dt^2} &\approx& \left ( \frac{2\pi}{P} \right)^3 a
\biggr [ \sin \left (\nu-\nu_0\right) + 
\nonumber \\
& & ~~~  2 e \left ( 11 \cos \nu \sin \left (\nu-\nu_0\right ) + 5 \sin \nu_0 \right ) \biggr ] 
\end{eqnarray}
so the jerk vanishes where
\begin{equation}
\sin \left ( \nu - \nu_0 \right ) = -10 e \sin \nu_0 + {\mathcal O} (e^2)
\end{equation}
To lowest order in the eccentricity, the radial velocity measurements
are equally sensitive to the timing at the primary ($\nu-\nu_0=0$) and
secondary transits ($\nu-\nu_0=\pi$), so we assume that the time
interval determined by fitting the radial velocity is given by the
average of the two intervals discussed earlier
\begin{equation}
\Delta t_{\rm RV} = \frac{1}{2} \left ( \Delta t_{0} + \Delta t_{\pi}
\right ) + {\mathcal O}(e^2) = P \left ( 1 - \frac{\delta
  \varpi}{2\pi} \right ) + {\mathcal O}(e^2).
\label{eq:4}
\end{equation}
The ``period'' obtained by fitting the radial velocity data differs
slightly by the period between periastrons or the period between
primary transits.  We confirmed this by generating radial velocity data with an
advancing periastron and fitting these data with purely Keplerian
radial velocity curves.  These simulations gave Eq.~\ref{eq:4} for
small eccentricities.

If we take the difference between the two observable quantities we get
\begin{equation}
\Delta t_{\rm RV} - \Delta t_0 = 2 e \cos \nu_0 \; \delta \varpi \; \frac{P}{2\pi}.
\end{equation}
The error in this quantity is given by
\begin{eqnarray}
\sigma_{\delta \varpi} &\approx& \left ( \sigma_{\Delta t_{\rm RV}}^2 + \sigma_0^2 \frac{12}{r_0 N^3} \right )^{1/2} \frac{2\pi}{P} \frac{1}{2e |\cos \nu_0|} \\
 &\approx& 10^{-5} 
 \left [\frac{\sigma_{\Delta t_{\rm RV}}^2}{(100~\rm{ms})^2}  + 0.12
 r_0^{-1} \frac{\sigma_0^2}{(10~{\rm s})^2} \left
 (\frac{N}{100}\right)^{-3}  \right ]^{1/2} \nonumber \\
& & ~~~ \times \frac{3~{\rm day}}{P} \frac{0.1}{e |\cos \nu_0|}
\end{eqnarray}
We see that the timing errors in the radial velocity measurements
dominate over the transit timing for a quoted precision in the
radial-velocity period of 100ms.  
Currently the best period estimates for hot Jupiters without transit
information are precise to 800ms \citep{2006ApJ...646..505B}, but
analysis of radial velocity
measurements over longer baselines would provide a more precise
estimate of this period.

Even without a detailed understanding of the relationship between the
velocity and transit timing, the radial velocity data is crucial to
convert a observed timing solution into a periastron shift by
determining the values of the eccentricity and the true anomaly at
transit, and in combination with the timing of either the primary or
secondary transit could yield hints of the periastron advance due to
other planets in the system.

As the orbit precesses, the radial velocities observed over a orbit
will also shift.  However, over the time required to detect a
Earth-like planet the orbit will only precess about an arcminute.  It
is difficult to imagine that radial velocity measurements will become
so sensitive as to characterize an orbit to the required precision of
$10^{-4}$.  If they did, one could probably detect the planet causing
the precession in the radial velocity data already.

\subsection{Rapid Precession}
\label{sec:rapid-precession}

If the transiting planet's orbital eccentricity goes to zero, the
precession rate of its pericenter longitude will formally go to
infinity, and Eq.~\ref{eq:3} misleadingly indicates that the
pericenter rate will be trivial to detect.  However, Eq.~\ref{eq:1} is
not correct in the case of rapid precession.

If the rate of the precession is a constant ($\delta \varpi$) per
radial orbit we have following equation for the true anomalies of the
transits
\begin{equation}
\nu \mod 2\pi = \left [ \nu_0 - \frac{ \delta \varpi}{P} t \right ] \mod 2\pi.
\label{eq:5}
\end{equation}
If we assume that the orbit is a precessing ellipse and that the
angular momentum of the observed planet is conserved, we
can use Kepler's equations to determine the time corresponding to each
true anomaly. To lowest order in the eccentricity we have
\begin{equation}
t = P \frac{\nu}{2\pi} - \frac{P}{2\pi} \frac{e^2}{4} \sin 2\nu
\end{equation}
If we first ignore the eccentricity, we find that the time between two
successive transits is given by the angular period
\begin{equation}
\Delta t^{(0)} = P \left ( 1 +  \frac{\delta\varpi}{2\pi} \right )^{-1}.
\end{equation}
and
\begin{equation}
\Delta \nu = \nu_2 - \nu_1 = 
-\delta\varpi \left ( 1 + \frac{\delta\varpi}{2\pi} \right )^{-1}
\end{equation}
Looking at Kepler's equation we find that the correction to this
quantity introduced by the eccentricity of the orbit is limited by
$e^2P/(4\pi)$.  We have
\begin{eqnarray}
\Delta t^{(2)} &=& \Delta t^{(0)}
- \frac{P}{2\pi} \frac{e^2}{4} \left ( \sin 2\nu_2 - \sin 2\nu_1 \right ) \\
 &=& \Delta t^{(0)}  \biggr \{ 1 
-  \nonumber \\ 
& & ~~~
\left ( 1 +  \frac{\delta\varpi}{2\pi} \right ) \frac{e^2}{4\pi} \left [ \sin \Delta\nu \cos \left ( 2\nu_1 + \Delta \nu \right ) \right ] \biggr  \}
\label{eq:7}
\end{eqnarray}
For a perturber in an elliptical orbit $\delta \varpi$ is inversely
proportional to the eccentricity of the observed planet
\citep{2000ssd..book.....M}; thus Eq.~\ref{eq:7} indicates that the
observable correction to the transit time is proportional to the
eccentricity, proving that the timing error becomes unobservable as $e$
tends to zero rather than diverging.

\section{Special Relativistic Corrections}
\label{sec:spec-relat-corr}

The foregoing analysis focused on the angles necessary for a transit
to occur.  It was essentially geometry with any kinematics.
Specifically it neglected the time for light to travel across the
system.  The variation in the distance of the planet and the star from
transit to transit as the orbit precesses would affect the times that
we observe the transits to occur.

The light travel time will affect the observed time
difference between the primary transit and the secondary transit that
immediately follows it \citep{2005ApJ...623L..45L},
\begin{eqnarray}
\left [ t_\pi - t_0 \right ]_{\rm obs}\!\!\!\! &=& \!\!\!\! \left [ {\mathcal M}(\nu_0+\pi) - {\mathcal M}(\nu_0)
  \right ] \frac{P}{2\pi} + \nonumber \\
& & 
\frac{1}{c} \left [  r(\nu_0) +  r(\nu_0+\pi)
  \right ]
\end{eqnarray}
where 
\begin{equation}
r(\nu) = \frac{a (1 - e^2)}{1+e\cos \nu}
\end{equation}
The light-travel time will cause the primary transit to appear to
occur about $ 20~\rm{s} \left ( \frac{P}{3~{\rm days}} \right )^{2/3}
\left ( \frac{M_*}{M_\odot} \right )^{1/3} $ earlier. The appearance
of the secondary transit will be delayed by similar interval.
This timing signature may be used to
constrain the physical size of the orbit; however, the difference in
the inter-transit interval due to geometry is on order of the period
of the orbit, a factor of $10^4$ larger, so the eccentricity of the
orbit must be known accurately for the relativistic corrections to be
useful.

Other quantities that we have examined are the the interval from
primary to primary transit and from secondary to secondary transit,
and the time derivative of these quantities.  If the orbit did not
precess the light travel time would not affect either of these
intervals, so we know that the relativistic correction to these
intervals will be proportional to the change in a angle of periastron
during an orbit.

We have
\begin{equation}
\left [ \Delta t_0 \right ]_{\rm obs} = P \left [ 1 - 
\left ( \frac{P}{2\pi} \left . \frac{d{\mathcal M}}{d\nu} \right |_0 - \frac{1}{c}  \left . \frac{dr}{d\nu} \right |_0 \right ) \dot \varpi  \right ]
\end{equation}
and
\begin{equation}
\left [ \Delta t_\pi \right ]_{\rm obs} = P \left [ 1 - 
\left ( \frac{P}{2\pi} \left . \frac{d{\mathcal M}}{d\nu} \right |_\pi + \frac{1}{c}  \left . \frac{dr}{d\nu} \right |_\pi \right ) \dot \varpi  \right ]
\end{equation}

The ratio of the two corrections to the interval between primary transits is
\begin{eqnarray}
\frac{\Delta t'_{0,{\rm rel}}}{\Delta t'_{0,{\rm ang}}} &=&
 \frac{\tan \nu_0}{\pi} \frac{a}{cP} \\ 
&\approx&
8 \times 10^{-5} \frac{\tan \nu_0}{\pi} \left ( \frac{M_*}{M_\odot} \frac{3~{\rm days}}{P} \right )^{1/3}.
\end{eqnarray}
The ratio is similar for the
secondary transits; therefore, as found by \citet{2005MNRAS.359..567A}
it is safe to ignore the relativistic corrections to the intervals
between two similar transits.

\section{Outlook}
\label{sec:outlook}

\subsection{Systems with a transiting planet}
\label{sec:syst-with-trans}

\begin{figure}
\includegraphics[width=3.4in]{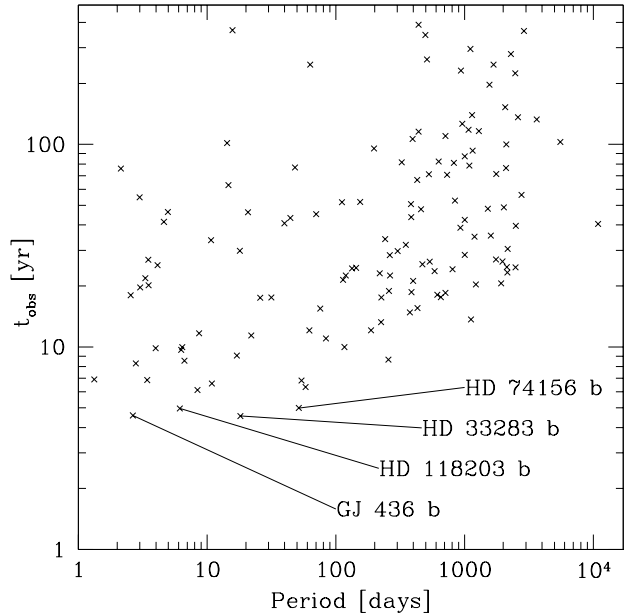}
\caption{The duration of the time series of primary and secondary
  transits (or primary transits and radial velocity timing) to yield a
  detectable $\delta\varpi$ of $10^{-7}$ for the planets in the
  catalogue of \citet{2006ApJ...646..505B}, assuming a ten-second
  error in the timing.  This is the sensitivity required to detect an
  Earth-mass planet at the three-sigma level within a factor of four
  in semimajor axis of the transiting planet, orbiting a solar mass
  star.  The required length of the time series increases as
  $\sigma_{\delta\varpi}^{-2/3}$.}
\label{fig:catalogue}
\end{figure}

Nearly two hundred extrasolar planets have been detected as of the
beginning of \citeyear{2006ApJ...646..505B}
\citep{2006ApJ...646..505B}\footnote{We used the updated catalogue at
  http://exoplanets.org.}.  The eccentricity and the longitude of
periastron have been measured for most of these planets;
unfortunately, for the few transiting planets, this information is
lacking, so we will use the orbital elements for all the nearby
exoplanets to calculate the sensitivity of timing measurements to
periastron advance and more importantly how long of a timing series
would be required to achieve a precision of $10^{-7}$ in the
measurement of $\delta\varpi$; this is sufficient to detect an
Earth-mass planet orbiting within a factor of four of the semimajor
axis of the transiting planet for a one-solar-mass host
star.  Fig.~\ref{fig:catalogue} shows that for about ten systems this
level of sensitivity could be achieved within a decade (these results
have assumed that the sampling rate $r$ is unity; determining the
optimal sampling rate and schedule considering observational
constraints is beyond the scope of this paper).

\begin{figure}
\includegraphics[width=3.4in]{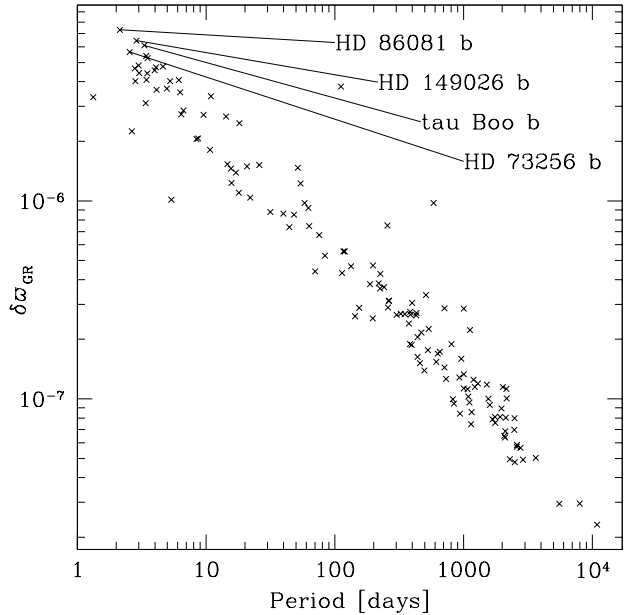}
\caption{The advance of the longitude of periastron per orbit for the
  planets of \citet{2006ApJ...646..505B} from general relativity. The
  four systems for which the relativistic advance is most rapid are
  HD~86081, HD~149026, tau~Boo, HD~73256.}
\label{fig:grres}
\end{figure}

To be certain of the presence of an unseen planet one has to be sure
that the periastron precession exceeds that expected from the star.
For this discussion we shall assume that the contribution due to
stellar oblateness is small.  According to Fig.~\ref{fig:wdot} even
for a three-day period, the expected contribution from oblateness is a
factor of 30$-$300 below the relativistic value.  This ratio increases
as $P^{2/3}$ so oblateness does not make the dominant contribution for
any of the systems in Fig.~\ref{fig:catalogue}.  On the other hand,
Fig.~\ref{fig:grres} shows the expected relativistic precession is
greater that $10^{-7}$ for most of the systems in the catalogue;
consequently, the mass of most of the host stars for these systems must
be determined to better than about a tenth of a solar mass.
Otherwise, the error in the determination of the mass of the host star
will dominate the statistical error in the timing.

\begin{figure}
\includegraphics[width=3.4in]{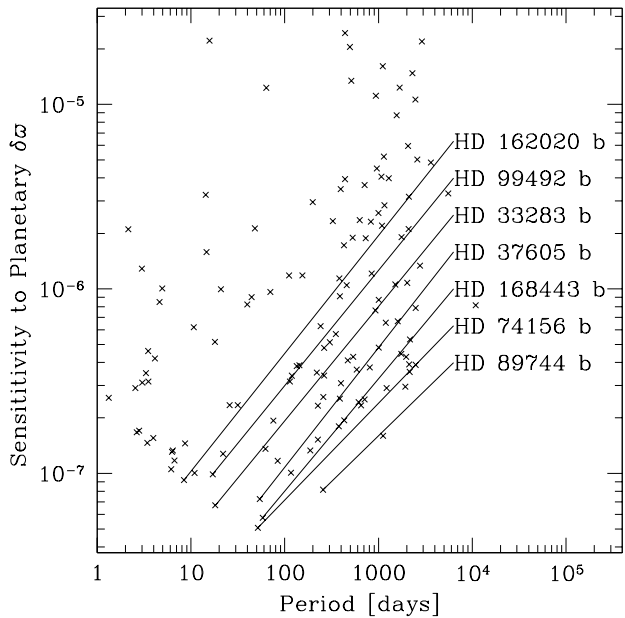}
\caption{The sensitivity to the advance of the longitude of periastron
  caused by additional planets (or stellar oblateness) in the
  planetary systems of \citet{2006ApJ...646..505B} assuming that the
  mass of the host star is known to $0.03\mbox{M}_\odot$ with a
  ten-year baseline of transit timing.  To detect an Earth-mass planet
  with an orbit up to a factor of four larger than the observed planet
  the sensitivity to the planetary contribution to the precession must
  be less than $10^{-7}$.}
\label{fig:plres}
\end{figure}

Fig.~\ref{fig:plres} shows the sensitivity to finding additional
planets in the transiting systems with a ten-year time series assuming
that the mass of the host star is known with a precision of
$0.03\mbox{M}_\odot$.
The properties of the transit itself, especially the shape of the
ingress and egress \citep{2005ApJ...623L..45L} and duration
\citep[e.g][]{2006astro.ph.12224W}, may constrain the mass of the star
further. Four objects stand out in
Figs.~\ref{fig:catalogue}--\ref{fig:plres}; GJ~436~b, HD~118203~b,
HD~33283~b and HD~74156~b require the shortest time series to achieve
a sensitivity of $10^{-7}$ in $\delta \varpi$.  These systems have
moderate eccentricities between 0.2 and 0.6.  Because of its short
2.6-day orbit, the orbital precession of GJ~436~b would be dominated
by relativistic effects with $(\delta \varpi)_\rmscr{GR} \approx 2.3
\times 10^{-6}$ assuming a mass of $0.41\mbox{M}_\odot$ for GJ~436
\citep{2006ApJ...646..505B}.  There are several planets in the
catalogue whose orbits will precess at a rate similar to GJ~436b due
to GR, but the geometry and eccentricity of the orbits are not as
favourable for detecting the precession by transit timing.
On the other hand, the maximum
sensitivity to unseen planets are found in the HD~74156 and HD~168443
systems because their relatively long orbits (52~days and 58~days)
reduces the expected contribution due to GR and stellar oblateness.
The minimum detectable value of $(\delta \varpi)_\rmscr{Planet}$ is
about $5\times 10^{-8}$, sufficient to detect an Earth-mass planet
with a period less than about 460 days
after a decade of observation.

Unfortunately, those planets that require the least time to constrain
additional bodies in the system have relatively long periods, on the
order of 50$-$100~days.  It is a factor of seven to ten times less
likely that planets with such long periods transit their star than the
hot Jupiters with three-day periods.  However, as more transiting
planets are discovered one would expect to find both planets with
longer periods and higher eccentricities than the current cohort of
transiting planets.  Indeed \citet{2006ApJ...646..505B} list several
planets that orbit their stars with periods of a few days in eccentric
orbits ($e\gtrsim 0.1$).  None of these systems have exhibited
transits, but it is only a matter of time before a transiting planet
with an eccentric orbit is found and these techniques may be brought
to bear.

\subsection{Systems with more than one observed planet}
\label{sec:systems-with-more}

A system that already has more than one planet detected provides a
chance to characterize an additional planet and verify that the stellar 
oblateness does not dominate over the planetary signal.  
Of course, each observed planet will induce a periastron precession in
the other, thus giving the planetary masses directly.  This must be
subtracted from the observed shifts along with the relativistic
shifts.  Because both planets orbit the same star the contribution due
to the stellar oblateness is proportional $P^{-7/3}$; consequently,
unless the residual periastron precession rate in each planet that
remains is proportional to $P^{-7/3}$, there must be other planets in
the system.

If one can argue from other data that the value of $J_2$ is small, one
can use the residual precession rate to find the mass of the unseen
planet and its semimajor axis.  With a single observed planet one can
only constrain the combination the determines $\delta \varpi$.  With
three observed planets, one could unravel $J_2$ and the properties of
an unseen planet or alternatively the properties of an unseen planet
and the possibility of further planets!

\subsection{Systems with fewer than one observed planet}
\label{sec:systems-with-fewer}

It may seem odd to suggest measuring the periastron advance in systems
without any planets yet discovered.  However, the techniques outlined
here could be used in eclipsing binary systems.  In single-lined
systems it could be used to constrain the total mass of the system
through the relativistic term.  In double-lined systems or if the
total mass of the system is known, measurements of the periastron
advance through timing of the radial velocity data and one or both
transits could be used to constrain the values of $J_2$ for the stars
or to look for planets in the systems.  One could search for planets
with the long-term timing measurements of eclipsing binaries that have
already been taken.

\section{Conclusions}
\label{sec:conclusions}

Accurate timing data (of pulsars) allowed the discovery of the first
extrasolar Earth-mass planets \citep{1992Natur.355..145W}.  Careful and accurate
timing of planet transits and radial velocity data is sensitive to
additional planets down to the mass of Earth and below.  The
combination of two sets of timing data provides much stronger
constraints on the presence of additional bodies in the system than
looking at the primary transits alone
\citep[c.f.][]{2002ApJ...564.1019M,2003sf2a.conf..149S,2005MNRAS.359..567A,2005Sci...307.1288H}.  
The timing signature of an Earth-mass planet is an induced shift in
the periastron of the orbits of the known planets.  There are
generally two dominant contributions to this shift: general relativity
and other planets.  The stellar oblateness can also contribute but
only competes with the other effects if the oblateness is nearly two
orders of magnitude larger than that of the Sun.  Consequently, if the
observed periastron shift exceeds the relativistic expectation, either
the system has additional planets or the parent star has an unusually
large oblateness.  Even the less likely possibility of a large
oblateness would give tantalizing hints to the origins of these
close-in extrasolar planets.  More likely would be the presence of a
Earth-mass or even a Mars-mass planet in a nearby orbit to the known
planet.

\section*{Acknowledgments}

We would like to thank Eric Agol, Josh Winn, Jaymie Matthews and the
referee for useful comments.  The Natural Sciences and Engineering
Research Council of Canada, Canadian Foundation for Innovation and the
British Columbia Knowledge Development Fund supported this work.
Correspondence and requests for materials should be addressed to
J.S.H. (heyl@phas.ubc.ca).  This research has made use of NASA's
Astrophysics Data System Bibliographic Services

\bibliographystyle{mn2e}
\bibliography{gr,planets}

\begin{thebibliography}{}

\bibitem[\protect\citeauthoryear{{Agol}, {Steffen}, {Sari} \&
  {Clarkson}}{{Agol} et~al.}{2005}]{2005MNRAS.359..567A}
{Agol} E.,  {Steffen} J.,  {Sari} R.,    {Clarkson} W.,  2005, \mnras, 359, 567

\bibitem[\protect\citeauthoryear{{Butler}, {Wright}, {Marcy}, {Fischer},
  {Vogt}, {Tinney}, {Jones}, {Carter}, {Johnson}, {McCarthy} \&
  {Penny}}{{Butler} et~al.}{2006}]{2006ApJ...646..505B}
{Butler} R.~P.,  {Wright} J.~T.,  {Marcy} G.~W.,  {Fischer} D.~A.,  {Vogt}
  S.~S.,  {Tinney} C.~G.,  {Jones} H.~R.~A.,  {Carter} B.~D.,  {Johnson} J.~A.,
   {McCarthy} C.,    {Penny} A.~J.,  2006, \apj, 646, 505

\bibitem[\protect\citeauthoryear{{Ford}, {Lystad} \& {Rasio}}{{Ford}
  et~al.}{2005}]{2005Natur.434..873F}
{Ford} E.~B.,  {Lystad} V.,    {Rasio} F.~A.,  2005, \nat, 434, 873

\bibitem[\protect\citeauthoryear{{Gladman}}{{Gladman}}{1993}]{Glad93}
{Gladman} B.,  1993, Icarus, 106, 247

\bibitem[\protect\citeauthoryear{{Hall}}{{Hall}}{1900}]{1900AJ.....20..185H}
{Hall} A.,  1900, \aj, 20, 185

\bibitem[\protect\citeauthoryear{{Holman} \& {Murray}}{{Holman} \&
  {Murray}}{2005}]{2005Sci...307.1288H}
{Holman} M.~J.,  {Murray} N.~W.,  2005, Science, 307, 1288

\bibitem[\protect\citeauthoryear{{Loeb}}{{Loeb}}{2005}]{2005ApJ...623L..45L}
{Loeb} A.,  2005, \apjl, 623, L45

\bibitem[\protect\citeauthoryear{{Mayor}, {Udry}, {Naef}, {Pepe}, {Queloz},
  {Santos} \& {Burnet}}{{Mayor} et~al.}{2004}]{2004A&A...415..391M}
{Mayor} M.,  {Udry} S.,  {Naef} D.,  {Pepe} F.,  {Queloz} D.,  {Santos} N.~C.,
    {Burnet} M.,  2004, \aap, 415, 391

\bibitem[\protect\citeauthoryear{{Miralda-Escud{\'e}}}{{Miralda-Escud{\'e}}}{2%
002}]{2002ApJ...564.1019M}
{Miralda-Escud{\'e}} J.,  2002, \apj, 564, 1019

\bibitem[\protect\citeauthoryear{Misner, Thorne \& Wheeler}{Misner
  et~al.}{1973}]{Misn73}
Misner C.,  Thorne K.~S.,    Wheeler J.~A.,  1973, Gravitation.
W. H. Freeman, San Francisco

\bibitem[\protect\citeauthoryear{{Murray} \& {Dermott}}{{Murray} \&
  {Dermott}}{2000}]{2000ssd..book.....M}
{Murray} C.~D.,  {Dermott} S.~F.,  2000, {Solar System Dynamics}.
Cambridge University Press

\bibitem[\protect\citeauthoryear{{Pireaux} \& {Rozelot}}{{Pireaux} \&
  {Rozelot}}{2003}]{2003Ap&SS.284.1159P}
{Pireaux} S.,  {Rozelot} J.-P.,  2003, \apss, 284, 1159

\bibitem[\protect\citeauthoryear{{Price} \& {Rush}}{{Price} \&
  {Rush}}{1979}]{1979AmJPh..47..531P}
{Price} M.~P.,  {Rush} W.~F.,  1979, American Journal of Physics, 47, 531

\bibitem[\protect\citeauthoryear{{Rasio}, {Nicholson}, {Shapiro} \&
  {Teukolsky}}{{Rasio} et~al.}{1992}]{1992Natur.355..325R}
{Rasio} F.~A.,  {Nicholson} P.~D.,  {Shapiro} S.~L.,    {Teukolsky} S.~A.,
  1992, \nat, 355, 325

\bibitem[\protect\citeauthoryear{{Schneider}}{{Schneider}}{2003}]{2003sf2a.con%
f..149S}
{Schneider} J.,  2003, in {Combes} F.,  {Barret} D.,  {Contini} T.,   {Pagani}
  L.,  eds, SF2A-2003: Semaine de l'Astrophysique Francaise {Multi-planet
  system detection by transits}.
pp 149--+

\bibitem[\protect\citeauthoryear{{Vogt}, {Butler}, {Marcy}, {Fischer}, {Henry},
  {Laughlin}, {Wright} \& {Johnson}}{{Vogt} et~al.}{2005}]{2005ApJ...632..638V}
{Vogt} S.~S.,  {Butler} R.~P.,  {Marcy} G.~W.,  {Fischer} D.~A.,  {Henry}
  G.~W.,  {Laughlin} G.,  {Wright} J.~T.,    {Johnson} J.~A.,  2005, \apj, 632,
  638

\bibitem[\protect\citeauthoryear{{Winn}, {Holman}, {Henry}, {Roussanova},
  {Enya}, {Yoshii}, {Shporer}, {Mazeh}, {Johnson}, {Narita} \& {Suto}}{{Winn}
  et~al.}{2006}]{2006astro.ph.12224W}
{Winn} J.~N.,  {Holman} M.~J.,  {Henry} G.~W.,  {Roussanova} A.,  {Enya} K.,
  {Yoshii} Y.,  {Shporer} A.,  {Mazeh} T.,  {Johnson} J.~A.,  {Narita} N.,
  {Suto} Y.,  2006, \aj, accepted (astro-ph/0612224)

\bibitem[\protect\citeauthoryear{{Winn}, {Noyes}, {Holman}, {Charbonneau},
  {Ohta}, {Taruya}, {Suto}, {Narita}, {Turner}, {Johnson}, {Marcy}, {Butler} \&
  {Vogt}}{{Winn} et~al.}{2005}]{2005ApJ...631.1215W}
{Winn} J.~N.,  {Noyes} R.~W.,  {Holman} M.~J.,  {Charbonneau} D.,  {Ohta} Y.,
  {Taruya} A.,  {Suto} Y.,  {Narita} N.,  {Turner} E.~L.,  {Johnson} J.~A.,
  {Marcy} G.~W.,  {Butler} R.~P.,    {Vogt} S.~S.,  2005, \apj, 631, 1215

\bibitem[\protect\citeauthoryear{{Wolszczan} \& {Frail}}{{Wolszczan} \&
  {Frail}}{1992}]{1992Natur.355..145W}
{Wolszczan} A.,  {Frail} D.~A.,  1992, \nat, 355, 145

\end{thebibliography}

\label{lastpage}

\end{document}